\documentstyle[eqsecnum,aps,epsfig]{revtex}

\def\Journal#1#2#3#4{{#1} {\bf #2}, #3 (#4)}

\def\NPA{{\em Nucl. Phys.} A}
\def\PLB{{\em Phys. Lett.} B}
\def\PRL{\em Phys. Rev. Lett.}
\def\PRC{{\em Phys. Rev.} C}

\newcommand{\bsigma}{\mbox{\boldmath $\sigma$}}
\newcommand{\btau}{\mbox{\boldmath $\tau$}}

\def\bq{\mbox{\boldmath$q$}}
\def\br{\mbox{\boldmath$r$}}
\newcommand{\bra}[1]{\langle #1 |}
\newcommand{\ket}[1]{| #1 \rangle}

\def\gnn{g'_{NN}}
\def\gnd{g'_{N \Delta}}
\def\gdd{g'_{\Delta\Delta}}

\begin{document}

\title{Pionic Modes Studied by Quasielastic $(\vec{p}, \vec{n})$ Reactions}

\author{M. Ichimura}

\address{Faculty of Computer and Information Sciences, Hosei University, \\
Kajino-cho 3-7-2, Koganei-shi, Tokyo 184-8584, Japan\\ 
E-mail: ichimura@k.hosei.ac.jp}

\author{K. Kawahigashi}

\address{Department of Information Sciences, Kanagawa University, \\
Hiratsuka-shi 259-1293, Japan\\
E-mail: ken@info.kanagawa-u.ac.jp}  


\maketitle

\begin{abstract}
\begin{quotation}
It has long been expected that the pionic modes show some collective phenomena
such as the pion condensation in the high density nuclear matter and 
its precursor phenomena in the ordinary nuclei. 
Here we show an evidence of the precursor observed 
in the isovector spin longitudinal cross sections $ID_q$ of the quasielastic 
$^{12}$C, $^{40}$Ca $(\vec{p}, \vec{n})$ reactions 
at $T_{\rm p} =$ 346 and 494MeV with the momentum transfer $q = $1.7fm$^{-1}$. 
Another aim of this report is to evaluate 
the Landau-Migdal parameters $\gnn$, $\gnd$ and $\gdd$ 
at the large momentum region from the above reactions. 
We obtained $\gnn \approx 0.6-0.7, \gnd \approx 0.3-0.4$. 
The results are consistent with those at the small momentum region, 
which are obtained from the Gamov-Teller strength distribution.
\end{quotation}
\end{abstract}

\section{Introduction}
I think that everybody agrees that the pions, predicted by Yukawa in 1935, 
still play a crucial role in nuclear physics. 
They couple with the nucleon in the form of $\btau\bsigma\cdot\bq$ 
where $\bq$ is the pion momentum. 
Therefore once the real or virtual pions are absorbed by the nucleus, 
they create the isovector spin longitudinal modes, i.e. the pionic modes. 

It has long been expected that the modes show some collective phenomena.
The most famous one is the pion condensation,\cite{Migdal} 
which may appear in high density nuclear matter such as the neutron star. 
Its appearance must have very important effects 
on the equation of state and the cooling processes of the neutron star.   

The condensation is considered not to realize in the normal nuclei 
because the critical density $\rho_{\rm c}$ of the phase transition 
has been evaluated to be much higher than the normal density $\rho_0$. 
However, we may expect to see some precursor phenomena 
even in the normal nuclei.\cite{EDTW,albe}
Among them, here,  we are interested in the enhancement of 
the isovector spin longitudinal response function 
\begin{eqnarray}
R_{\rm L}(q,\omega) = \sum_{n}|\bra{n}\sum_{i} \tau_{i}^{a}
(\bsigma_{i} \cdot \hat{\bq})e^{i\bq\cdot\br_{i}}\ket{0}|^{2}
\delta(\omega-(E_{n}-E_{0})) 
\end{eqnarray}
around the quasielasitc region, where $\bq$ is the momentum transfer, 
$\hat{\bq}$ is its unit vector, $\omega$ is the energy transfer and 
$E_n$ is the intrinsic energy of the $n$-th nuclear state 
(0 denoting the ground state). 
The superscript $a$ denotes the isospin component, $a = +, 0, -$, 
where $a = -$ corresponds to those obtained by the $(p, n)$ reaction.
To make the notation simple, 
we suppress the superscript $a$ in the r.h.s. and below.

I would like to make the following two comments. 
The first is that the real or virtual photon 
couples with the nucleon in the form of $\btau\bsigma\times\bq$ 
for the spin-isospin channel, and 
thus creates only the isovector spin transverse modes 
but not the spin longitudinal modes. 
Therefore the reliable probe $(e, e')$ is not suitable 
to investigate the pionic modes, but  
it fits to see the isovector spin transverse response function
\begin{eqnarray}
R_{\rm T}(q,\omega) = \frac{1}{2}\sum_{n}\sum_{\mu}|\bra{n}\sum_{i} \tau_{i}^{a} 
(\bsigma_{i} \times \hat{\bq})_{\mu} e^{i\bq\cdot\br_{i}}\ket{0}|^{2}
\delta(\omega-(E_{n}-E_{0}))    
\end{eqnarray}
In this report I will not touch the spin transverse part 
because of lack of time. 

Another point is that 
the momentum $\bq$ is not a good quantum number in the actual nuclei, 
because they are of finite size. 
Therefore the spin longitudinal and the spin transverse modes are mixed 
in a certain discrete state. 
Each mode is selected by the probe used. Here we are concerned only with 
the gross structure of the energy spectra around the quasielastic peak.

\section{Origin of the collectivity\ --Effective interactions}

The collectivity of the pionic modes is considered to come from 
(1) the attraction of OPEP and (2) the coupling with the $\Delta$-hole states

To evaluate the collectivity we need the effective interactions 
for the ph-ph, ph-$\Delta$h and $\Delta$h$-\Delta$h couplings. 
The ph-ph part of the spin-isospin channel is written as 
\begin{equation}
V(q,\omega) = (\btau_1\cdot\btau_2)
    [V_{\rm L}(q,\omega)(\bsigma_1\cdot\hat{\bq})(\bsigma_2\cdot\hat{\bq}) 
 + V_{\rm T}(q,\omega)(\bsigma_1\times\hat{\bq})(\bsigma_2\times\hat{\bq})]
\end{equation}
where the $V_{\rm L}$ and $V_{\rm T}$ parts are 
the spin longitudinal and the spin transverse interactions, respectively.
The expression is equivalent to the familiar form, 
the sum of the central and the tensor parts.
Extension to the ph-$\Delta$h and $\Delta$h$-\Delta$h interactions 
are straightforward.

They are often represented by the $\pi + \rho + g'$ model, 
which expresses the interaction as 
the one pion exchange plus the one $\rho$-meson exchange plus 
the contact interaction specified by the three Landau-Migdal parameters, 
$\gnn$, $\gnd$ and $\gdd$. 
In this model $V_{\rm L}$ and $V_{\rm T}$ are expressed as
\begin{eqnarray}
V_{\rm L}(q,\omega) &=& \frac{f_{\pi NN}^2}{m_\pi^2}
                        \left(g' + \frac{q^2}{\omega^2-q^2-m_\pi^2}
                        \Gamma_\pi^2(q,\omega)\right), \label{eq:VL} \\
V_{\rm T}(q,\omega) &=& \frac{f_{\pi NN}^2}{m_\pi^2}
                        \left(g' + \frac{q^2}{\omega^2-q^2-m_\rho^2}
                        C_{\rho}\Gamma_\rho^2(q,\omega)\right) \label{eq:VT}.
\end{eqnarray}
where $C_{\rho}$ is the coupling ratio between the $\pi$ and $\rho$ exchange 
and $\Gamma_{\pi(\rho)}$ are the coupling form factor 
of the $\pi(\rho) NN$ vertex.
Their $q$ dependence are shown in Fig. 1 for $\omega =$ 80MeV and $g' = 0.6$. 
The genuine one pion exchange part 
$V_{\pi}$ ($g'= 0$) is also shown (dashed line). 
We see that $g'$ represents the strength at $q = 0$ 
and determines 
above what $q$ the interaction $V_{\rm L}$ becomes negative and 
how attractive it is at large $q$, 
which are crucial for the pionic collectivity.
Therefore it is one of the main aims of this report to estimate the $g'$'s. 

\begin{figure}[ht]
\begin{center}
\epsfxsize=17pc 
\epsfbox{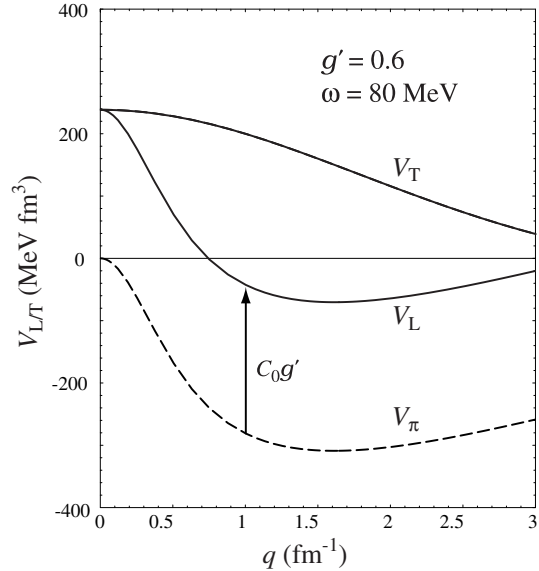} 
\end{center}
\caption{Effective interaction. $C_0 = f_{\pi NN}^2/m_\pi^2$.
\label{fig:Veff}}
\end{figure}

Recently Suzuki and Sakai\cite{suzuki} phenomenologically estimated 
them from the energy of the giant Gamov-Teller (GT) resonance and 
the quenching factor $Q$ for the GT sum rule value 
observed by Wakasa {\em et al.},\cite{wak_zr} where $Q$ is defined as 
\begin{equation}
Q = \frac{S_{\beta^-}^{\rm exp} - S_{\beta^+}^{\rm exp}}{3(N-Z)} \quad
{\rm with} \quad
S_{\beta^{\mp}} = \sum_n |\bra{n}\tau^{\pm}\bsigma\ket{0}|^2.
\end{equation}
Using the Fermi gas model and the sum rule technique, they obtained 
\begin{equation}
0.58 < g'_{NN} < 0.59, \quad 0.18 < g'_{N \Delta} < 0.23 \qquad 
{\rm if} \quad \gdd \le 1.0,
\end{equation}
Arima {\em et al.}\cite{arima_00} reported 
that the finiteness increases $\gnd$ by about 0.1.

This is the estimation at $q \approx 0$. Then what happens at large $q$ ?

\section{Experimental method to study the pionic modes}

Since the reliable probe $(e, e')$ does not work 
for study of the spin longitudinal responses, 
we have inevitably to utilize the hadronic probes 
though their reaction mechanism is more complex.
The best one could be the $(\vec{p}, \vec{n})$ reactions, 
for which we need the polarized proton beam with high intensity and 
the neutron polarimeter with high efficiency. 

Due to these difficulties, complete measurement of 
the polarization transfer coefficients $D_{ij} 
(D_{SS'},\,D_{NN'},\,D_{LL'},\,D_{SL'},\,D_{LS'})$
 of the $(\vec{p}, \vec{n})$ reactions became possible 
 just in 1990's.\cite{terry,wakold} 
The coefficient $D_{ij}$ represents the transition probability 
from the proton with the spin polarization in the direction $i$ 
to the neutron with the spin polarization in the direction $j$. 
The directions are specified in Fig. 2.

\begin{figure}[ht]
\begin{center}
\epsfxsize=16pc 
\epsfbox{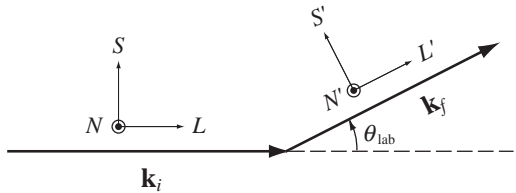} 
\end{center}
\caption{Spin direction assignment \label{fig:scatter}}
\end{figure}

To extract the spin response functions, Bleszynski {\em et al.}\cite{blesz}
introduced the spin longitudinal cross section $ID_{q}$ and 
the spin transverse cross section $ID_{p}$, etc., 
which are defined by the linear combinations of $D_{ij}$ times 
the unpolarized cross section $I$. 
The naming comes from the fact that 
$ID_{q(p)}$ is proportional to $R_{\rm L(T)}$ as 
\begin{eqnarray}
ID_q(q, \omega) &=& K (N_{\rm eff}/N)I^{NN}D_q^{NN}R_{\rm L}(q, \omega) \\
ID_p(q, \omega) &=& K (N_{\rm eff}/N)I^{NN}D_p^{NN}R_{\rm T}(q, \omega) 
\end{eqnarray}
if the plane wave impulse approximation (PWIA) 
with the $N_{\rm eff}$ prescription works.\cite{blesz,IK92} 
Here $K$ is the kinematical factor, 
$N_{\rm eff}$ is the number of neutrons which participate the reaction and 
$I^{NN}D_i^{NN}$ is the corresponding quantity of the $NN$ scattering. 

In this report we concentrate on only the pionic modes, 
consequently only the spin longitudinal cross sections $ID_q$.

\section{Theoretical approach to the spin longitudinal response}

We must fully take into account 
the finite size feature of the nucleus, 
since it mixes up the spin longitudinal and the spin transverse modes.
On top of that we are interested in the gross collective nature 
in the continuum. 
Therefore we employed the continuum random phase approximation (RPA) method 
in the angular momentum representation. 

In preparation of the single particle states and 
the single particle Green's functions, we use 
a radial dependent effective mass $m^{*}(r)$, 
and treat the spreading widths of the holes and the particles 
by the complex binding energy and the complex potential, 
respectively.\cite{KNII01}

The $\Delta$-hole configurations are included in the continuum RPA formalism 
and the $\pi + \rho + g'$ model interaction is utilized. 
Most of previous analyses were carried out in the approximation 
$\gnn = \gnd =\gdd$ , which is called the universality ansatz. 
We remove this ansatz and treat these $g'$'s independently. 

In hadronic reactions the distortion and absorption must be taken into account. 
Therefore we developed a formalism\cite{KNII01} 
of the distorted wave impulse approximation (DWIA) 
for the continuum final states incorporating with the continuum RPA. 

\section{Numerical analysis of 
$^{12}$C, $^{40}$Ca $(\vec{p}, \vec{n})$ reactions}

Using the above DWIA + continuum RPA method, 
we analyzed the $^{12}$C, $^{40}$Ca$(\vec{p}, \vec{n})$ reactions 
at $T_{\rm p} =$ 346MeV and $\theta = 22^{\circ}$ 
by Wakasa {\em et al.}\cite{wakold,waknew} 
and at $T_{\rm p} =$494MeV and $\theta = 18^{\circ}$ 
by Taddeucci {\em et al}.\cite{terry} 
In the both cases the transferred momentum $q$ is about 1.7fm$^{-1}$ 
for $\omega < 120$MeV.

\begin{figure}[ht]
\begin{center}
\epsfxsize=16pc 
\epsfbox{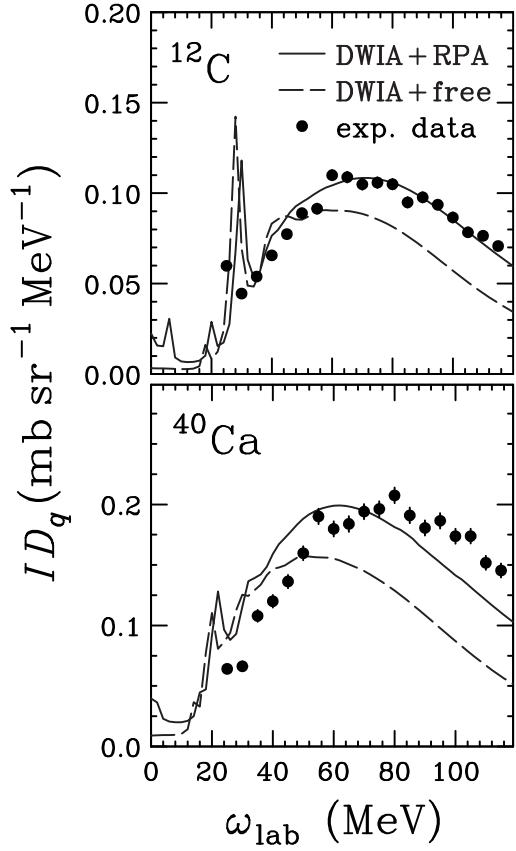} 
\end{center}
\caption{The spin longitudinal cross section $ID_q$ of 
$^{12}$C,$^{40}$Ca$(\vec{p}, \vec{n})$ at $T_{\rm p}$ = 346MeV 
and $\theta = 22^{\circ}$. 
In RPA $(\gnn, \gnd, \gdd) = (0.7, 0.3, 0.5)$ and $m^*(0) = 0.7m$ are used.  
\label{fig:ID-q-346}}
\end{figure}

\begin{figure}[ht]
\begin{center}
\epsfxsize=15.5pc 
\epsfbox{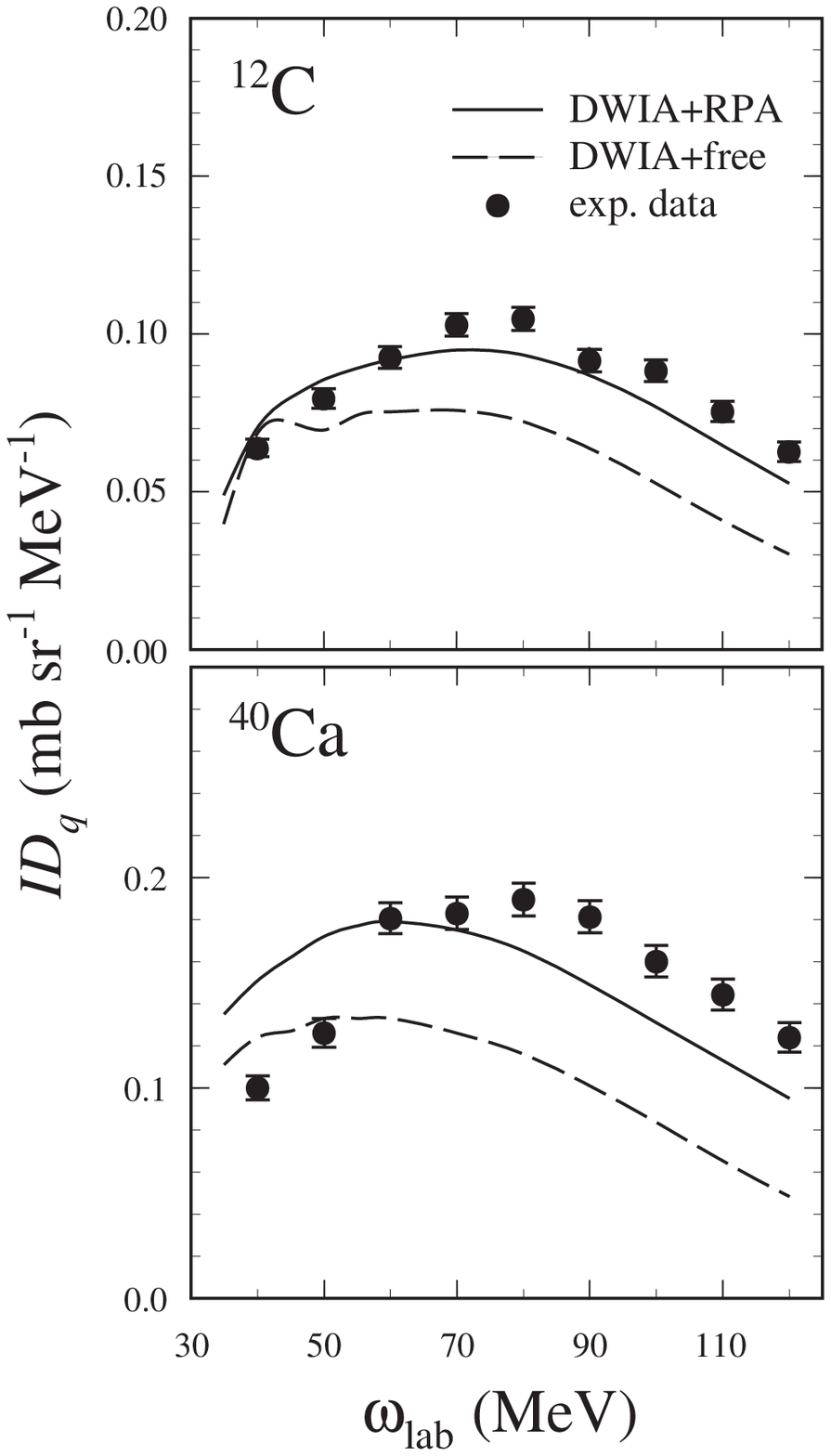} 
\end{center}
\caption{$ID_q$ of $^{12}$C,$^{40}$Ca$(\vec{p}, \vec{n})$ at $T_{\rm p}$ = 494MeV 
and $\theta = 18^{\circ}$. 
In RPA,  $(\gnn, \gnd, \gdd) = (0.6, 0.3, 0.5)$ and $m^*(0) = 0.7m$ are used.  
\label{fig:ID-q-494}}
\end{figure}

We treated the Landau-Migdal parameters $\gnn, \gnd$ and 
the effective mass at the nuclear center $m^*(r=0)$ as adjustable parameters.
The calculated spin longitudinal cross sections $ID_q$'s 
with and without the RPA correlations are compared 
to the experimental results 
in Fig. \ref{fig:ID-q-346} for the $T_{\rm p}$ = 346MeV cases and 
Fig. \ref{fig:ID-q-494} for the $T_{\rm p} =$494MeV cases.
As for the exprimental results of the $T_{\rm p}$ = 346MeV cases, 
we quoted the latest results\cite{waknew} 
instead of the published ones.\cite{wakold}

We could reproduce $ID_q$ reasonably well if we include the RPA correlation 
with the parameters $\gnn = 0.6-0.7$, $\gnd = 0.3 - 0.4$ and $m^*(0) = 0.7m$. 
We fixed $\gdd = 0.5$ but its dependence is very weak.
Comparing to the results without the correlation 
(denoted by "free" in the figures), we clearly see 
the enhancement of the spin longitudinal cross sections. 
Thus we first found the precursor phenomena of the pion condensation.

\section{Conclusion}
We analyzed the quasielastic $(\vec{p}, \vec{n})$ reactions 
by the DWIA + continuum RPA method. 
Adjusting $\gnn,\  \gnd$ and $m^*(0)$, 
we could reproduce the observed spin longitudinal cross section $ID_q$. 

The analysis shows in the first time an experimental evidence 
of the precursor phenomena of the pion condensation,
i.e. the enhancement of the spin longitudinal response function $R_{\rm L}$.

Our estimation of $g'$'s at $q = 1.7$fm$^{-1}$ from $ID_q$ gives
$\gnn = 0.6-0.7, \gnd = 0.3 - 0.4$, 
which are close to those at $q \approx 0$ estimated 
from the GT strength distribution.
This means that the $\pi + \rho + g'$ model with constant $g'$'s works well 
for wide range of $q$.

Our findings imply that the various old calculations of the critical density 
$\rho_{\rm c}$ of the pion condensation based on the universality ansatz 
should be redone. It may give a great modification 
on the equation of state and the scenario of the cooling of the neutron star. 

I did not touch the spin transverse response functions 
due to shortage of the time, 
but I must mention at the end that the observed 
$R_{\rm T}$'s by $(p, n)$ are much larger than 
those obtained by $(e, e')$ and this contradiction must be solved 
before reaching the definite conclusion.

\section*{Acknowledgments}
We would like to express our gratitude to Dr.\  T.\  Wakasa 
for providing us with new experimental results before publication. 
This work was supported in part by the Grants-in-Aid 
for Scientific Research Nos.~02640215, 05640328 and 12640294 
of the Ministry of Education, Science, Sports and Culture of Japan.

\end{document}